\documentclass[aps,prd,twocolumn,groupedaddress]{revtex4-2}

\usepackage{graphicx,times}
\usepackage{natbib}
\usepackage{amssymb,amsmath}
\usepackage{longtable}
\usepackage{aas_macros}

\begin{document}

\title{Estimation of black hole spins in low-mass AGNs and comparison with other types of AGNs}

\author{M.Yu. Piotrovich}
\email[]{mpiotrovich@mail.ru}
\author{S.D. Buliga}
\author{T.M. Natsvlishvili}

\affiliation{Central astronomical observatory at Pulkovo, St.-Petersburg, 196140, Russia \\  
\\ Accepted for publication in Physical Review D}

\date{\today}

\begin{abstract}
We estimated the spins of a sample of 58 low-mass AGNs. Analysis of the obtained spins showed that they decrease with increasing SMBH mass, leading us to hypothesize that mergers and/or chaotic accretion are the primary mechanisms for mass growth. In this regard, we proposed a more general hypothesis about the evolution of AGNs. We assume that early low-mass SMBHs have high spins, then, during their evolution, the spins initially decrease and then begin to increase, with the rate of increase gradually slowing.
\end{abstract}


\maketitle

\section{Introduction}

Supermassive black holes (SMBHs) are believed to reside in the centers of most galaxies and play a fundamental role in their formation and evolution. In active galactic nuclei (AGNs), accretion of matter onto the central black hole releases enormous amounts of energy across the electromagnetic spectrum. While the mass of the black hole is commonly considered the primary parameter governing AGN properties, the dimensionless spin parameter $a$ is increasingly recognized as a key factor influencing accretion efficiency, jet production, and feedback processes. Therefore, measurements of black hole spin are essential for understanding the physics of accretion and the co-evolution of black holes and their host galaxies.

Although actual measurements of a black hole's rotation rate are challenging, significant progress has been made and there are now many methods available to probe the rotation of an accreting black hole. For example, methods have been developed to determine the spin based on the analysis of the Fe~K$\alpha$ X-ray line profiles and the analysis of the spectral emission energy distribution (SED) of the accretion disk \citep{reynolds14,brenneman13,done13,capellupo16,middleton16,mallick22}. However, these methods require high-quality spectroscopic and photometric data. Moreover, different spectral processing methods often yield significantly different results even for the same object. This is due to the model dependence of the methods used. These circumstances stimulated the search for other approaches to constraining the spin magnitude.

Beginning with the pioneering works of \citet{blandford77} and \citet{blandford82}, it was shown that spin must play an important role in the energy release process. The Blandford–Znajek and Blandford–Paine mechanisms are invoked to explain the energetics of jets—the jet-like outflows observed in various types of AGN. The BZ mechanism describes the contribution to the jet energy due to the extraction of rotational energy from the black hole. In the BP mechanism, the jet is created by outflowing matter from the accretion disk. The BZ and BP mechanisms can act jointly. This assumption underlies the so-called hybrid models \citep{meier99,garofalo09,garofalo10,daly09}. The relationship between jet power and spin opens another approach to obtaining constraints on the spin \citep{daly11,daly14,gnedin14,piotrovich16a}.

Furthermore, the thermal continuum fitting (CF) method is used to estimate the spin. It is based on the fact that the black hole's rotation will affect the temperature of the inner disk, and gradual accretion onto a rapidly rotating black hole leads to the highest temperatures \citep{zhang97,mcclintock14}. Within a specific accretion disk model, this can be quantified and turned into a precision tool for determining the black hole's rotation.

Other methods for determining AGN spins are known from the literature: spin determination from quasi-periodic oscillations (QPO); using microlensed quasars; and spin determination from direct images. And, of course, gravitational wave observations have provided fundamentally new opportunities for studying black hole physics, including AGN spins. These methods are described in more detail in the review of \citet{reynolds21}.

In this paper, we used the radiative efficiency method to determine AGN spins. The radiative efficiency of the AGN, $\varepsilon(a)$, determines the process of conversion of gravitational energy into radiation and strongly depends on the spin of the SMBH \citep{novikov73,davis06,krolik07,krolik07b,schnittman16}. The method we used is described in more detail in Section III.

Low-mass active galactic nuclei (LMAGNs), hosting black holes with masses typically below $\sim 2\times 10^{6} M_{\odot}$ \citep{greene07,gultekin14}, provide a particularly valuable laboratory for studying black hole growth in relatively unevolved systems \citep{greene06,greene07,greene12,greene20,koliopanos17,gultekin14,volonteri21,trinca22,kocevski23,pucha25,grishin25}. These objects may represent an important population linking intermediate-mass black holes and classical luminous AGNs powered by more massive SMBHs. It should be noted that they also can form in the direct collapse scenario. However, despite their significance, the spin properties of black holes in LMAGNs remain poorly constrained. This is primarily due to observational challenges, including their comparatively low luminosities, complex spectral features, and the limited sensitivity of current observations.

In this work, we estimated the spins of black holes in low-mass active galactic nuclei using currently available observational data. The results are relevant for constraining black hole growth histories, testing accretion disk theory, and improving our understanding of the role of spin in AGN feedback processes.

\section{Selection of low-mass AGNs and their analysis}

\begin{figure}[ht!]
\includegraphics[bb= 30 10 715 530, clip, width=\linewidth]{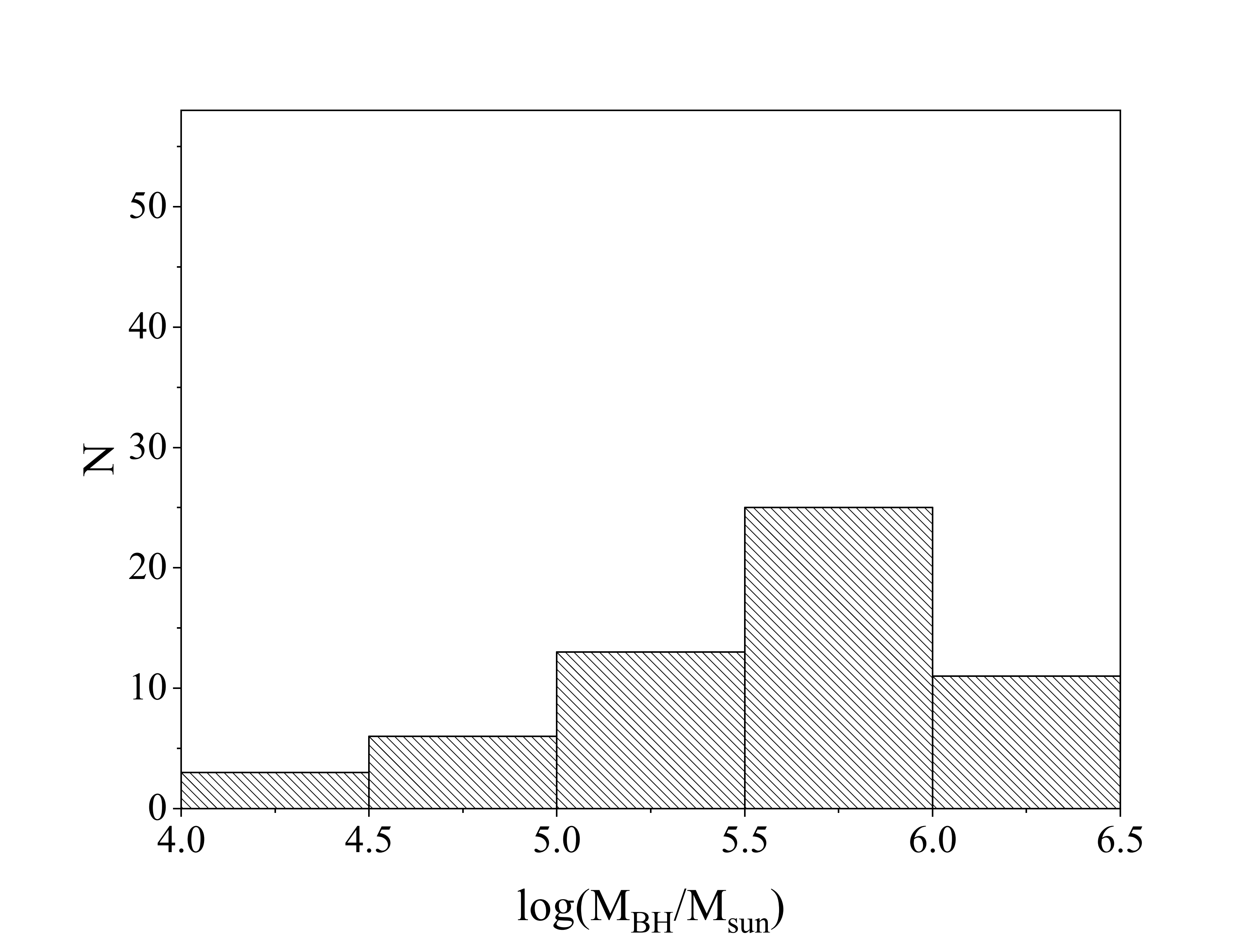}
\caption{Distribution of the SMBH mass in our low-mass AGN sample.
\label{fig:hist_MBH}}
\end{figure}

\begin{figure}[ht!]
\includegraphics[bb= 30 10 715 530, clip, width=\linewidth]{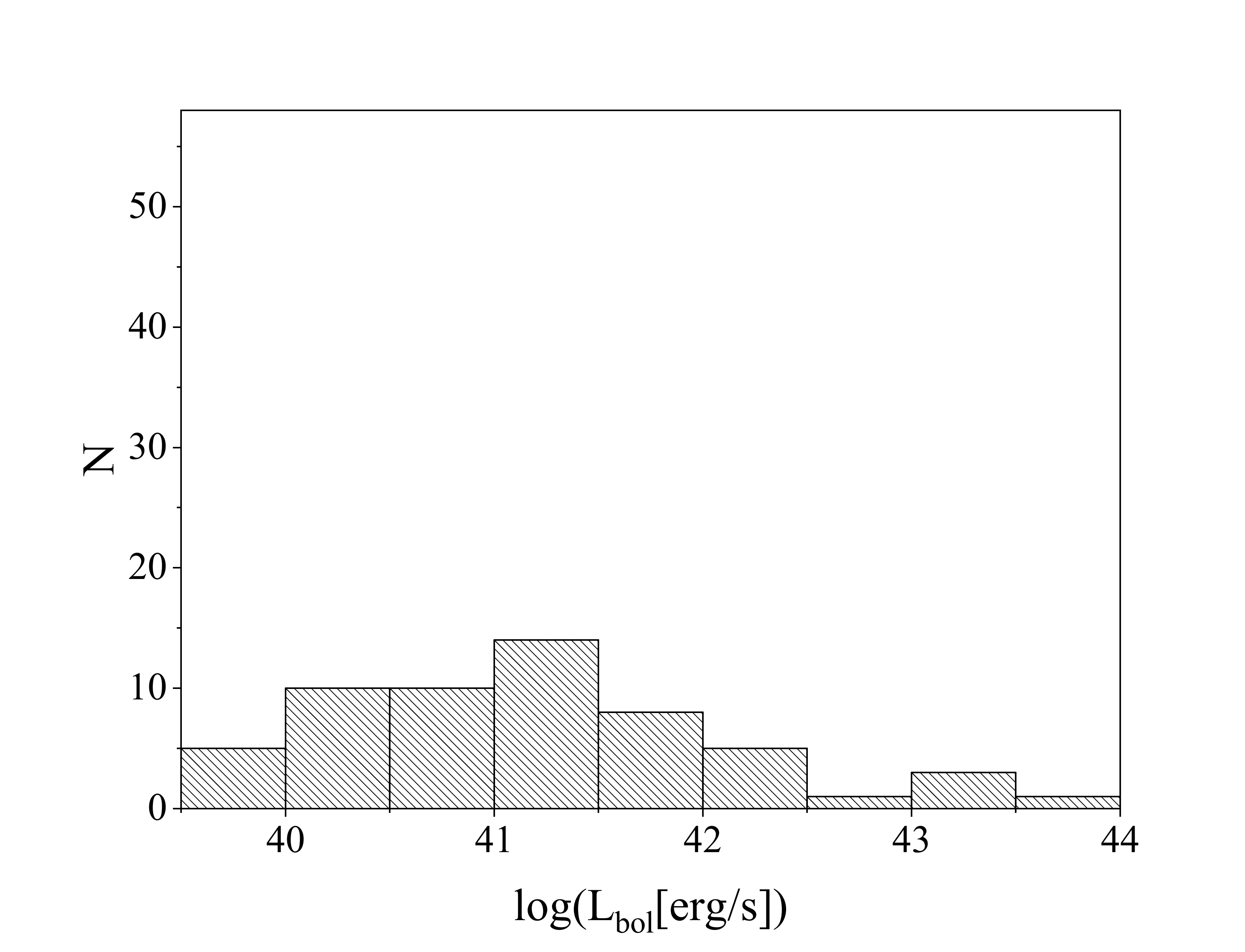}
\caption{Distribution of the bolometric luminosity in our low-mass AGN sample.
\label{fig:hist_Lbol}}
\end{figure}

\begin{figure}[ht!]
\includegraphics[bb= 30 10 715 530, clip, width=\linewidth]{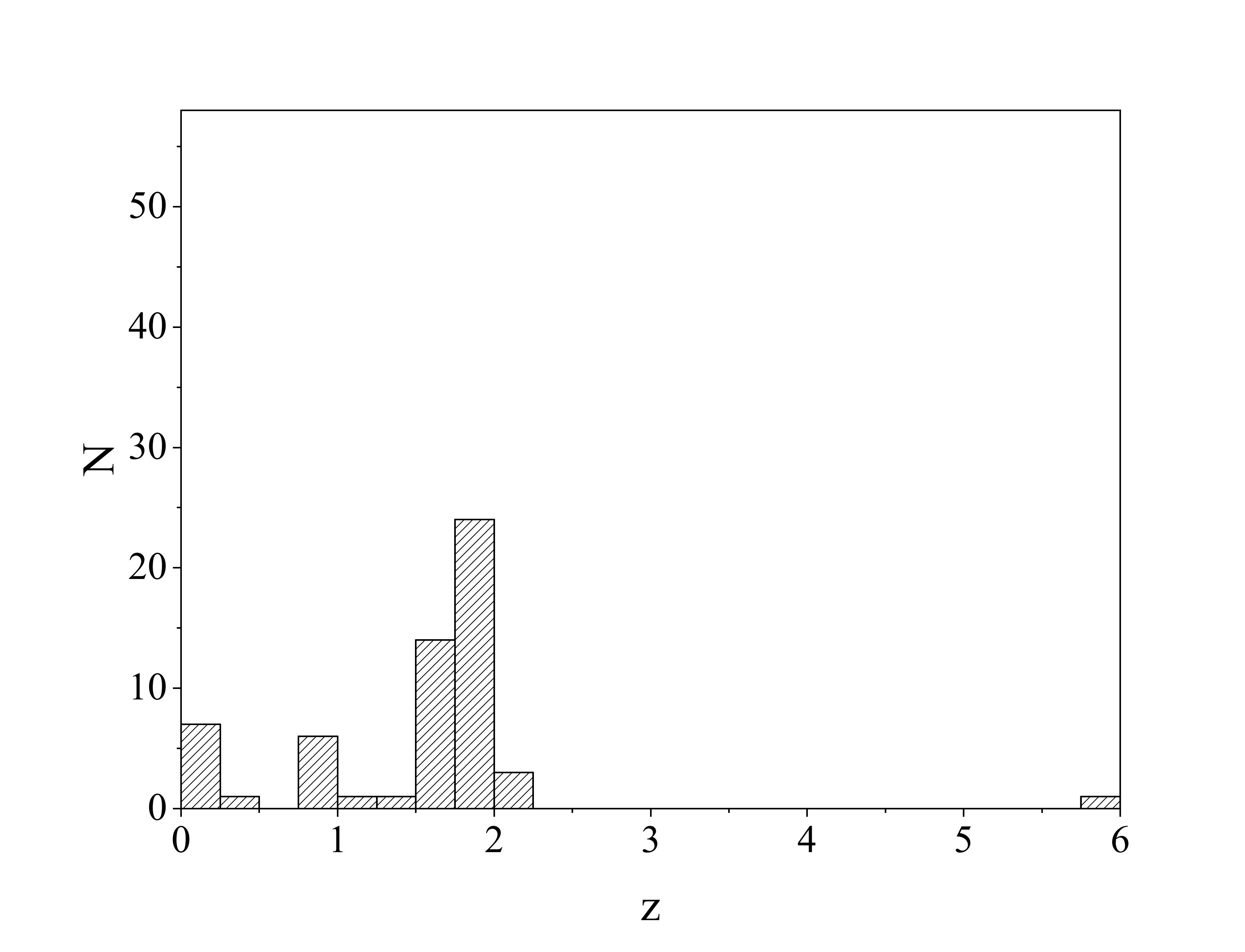}
\caption{Distribution of the cosmological redshift in our low-mass AGN sample.
\label{fig:hist_z}}
\end{figure}

\begin{figure}[ht!]
\includegraphics[bb= 30 10 715 530, clip, width=\linewidth]{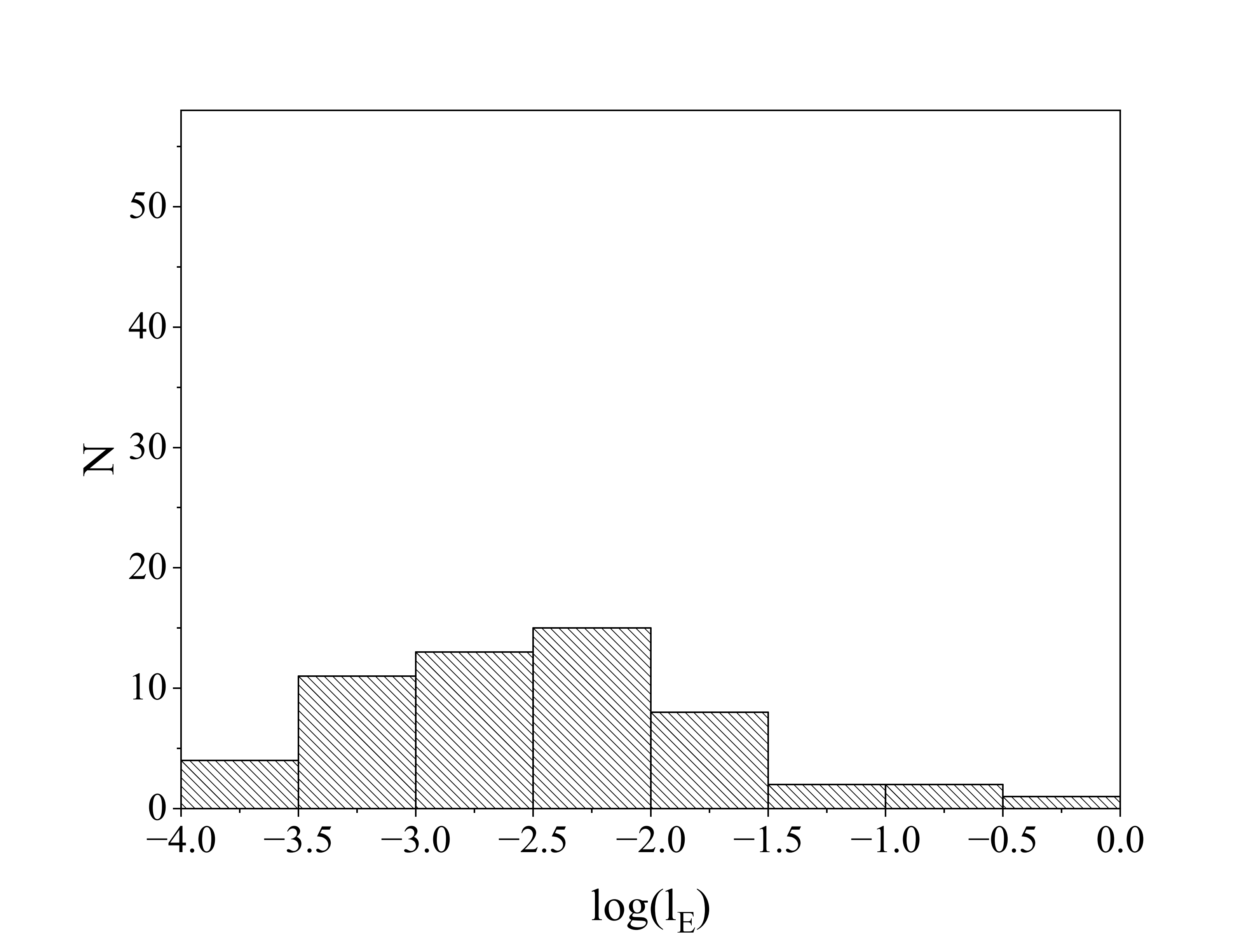}
\caption{Distribution of the Eddington ratio in our low-mass AGN sample.
\label{fig:hist_lE}}
\end{figure}

\begin{figure}[ht!]
\centering
\includegraphics[bb= 30 10 715 535, clip, width=\linewidth]{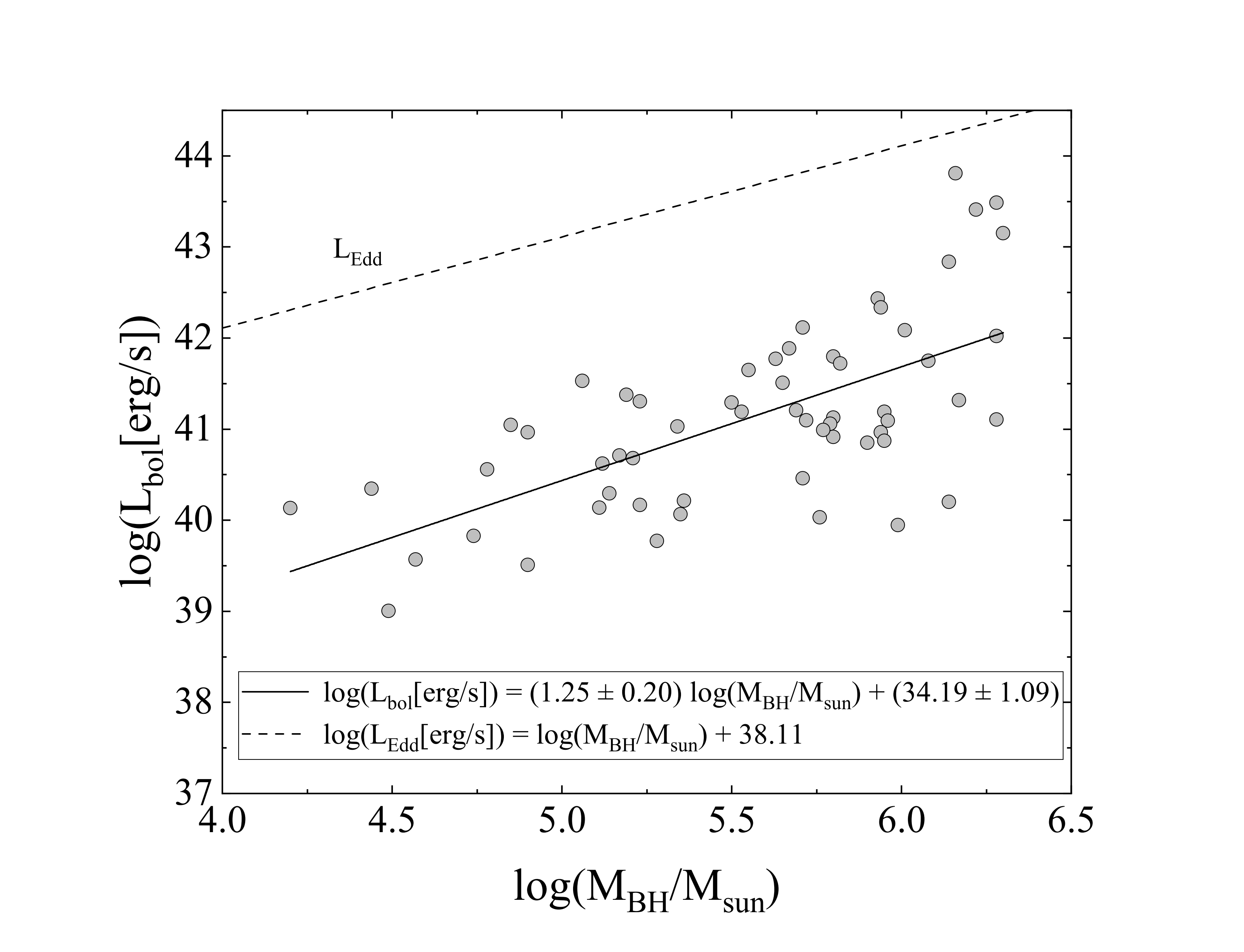}
\caption{Dependency of the bolometric luminosity on the SMBH mass in our low-mass AGN sample. $\log{L_{\rm Edd}{\rm [erg/s]}} = \log(M_{\rm BH} / M_\odot) + 38.11$ is the Eddington luminosity.
\label{fig:Lbol_MBH}}
\end{figure}

As a source, we used the catalog from \citet{wu22}. It contains properties we need for 750414 broad-line quasars from Sloan Digital Sky Survey Data Release 16 quasar catalog (SDSS DR16Q).

In this work, we adopt the cosmological model used in \citet{wu22} catalog: a flat ${\rm \Lambda}$CDM cosmology with $\Omega_{\rm \Lambda} = 0.7$, $\Omega_M = 0.3$, and $H_0 = 70$km s$^{-1}$ Mpc$^{-1}$.

This catalogue contained 63 objects with a mass $\leq2\times10^6M_\odot$. For the object SDSS~J010407.53+085816.8, several significantly different redshift values are given in different catalogs, so we decided to exclude it from the sample. The object SDSS~J133816.45+371639.8 may not be a quasar at all (according to Simbad data), so we also exclude it. The objects SDSS~J025023.55-041050.8 and SDSS~J010137.75+195548.6 stand out very much from the sample, having an anomalously low bolometric luminosity ($\sim 10^{37}$ erg/s), so we also decided to exclude them from consideration. The object SDSS~J013803.07+000939.6 has a super-Eddington accretion rate, and since our calculation method uses the Shakura-Sunyaev accretion disk model \citep{shakura73}, we also discard it. This leaves us with 58 objects.

Next, we performed a statistical analysis of the initial sample.

Figs.\ref{fig:hist_MBH}-\ref{fig:hist_lE} show the distribution of the SMBH mass, the bolometric luminosity, the cosmological redshift and the Eddington ratio in our sample. All distributions except the redshift distribution appear close to log-normal. The redshift distribution appears close to normal (similar to other types of AGN) in the range $1.0 \leq z \leq 2.5$ with peak at $1.8 < z < 2.0$, but for $z < 1.0$ and $z > 2.5$ (one object with $z = 5.8$), the objects do not fit the normal distribution. The small number of objects and the selection bias are likely at play here. At close distances, objects are few, while at greater distances, observing objects becomes more difficult and we see fewer of them. We will take this fact into account in further calculations. The peak in mass distribution falls on the region $5.5 < M_{\rm BH} < 6.0$, the peak in the bolometric luminosity distribution is at $41.0 < \log{L_{\rm bol}{\rm [erg/s]}} < 41.5$, the peak in Eddington ratio distribution is at $-2.5 < \log{l_{\rm E}} < -2.0$. We note in general the rather low values of the Eddington ratio. This may be due, for example, to the fact that these AGNs are located in dwarf galaxies and have low accretion rate \citep{yuan14b}. It should be noted that this also may be caused by the fact that the emission of the nucleus is buried in the host starlight.

Fig.\ref{fig:Lbol_MBH} shows the dependency of the bolometric luminosity on the SMBH mass and in our sample. A strong correlation is observed between luminosity and mass (Pearson correlation coefficient is 0.65). Linear fitting gives us:
\begin{multline}
\log(L_{\rm bol}{\rm [erg/s]}) =\\
= (1.25 \pm 0.20) \log(M_{\rm BH}/M_\odot) + (34.19 \pm 1.09).
\label{eq:Lbol_MBH}
\end{multline}
A slope of $\sim 1.25$ is quite large, for comparison, for sample of red quasars we had $\sim 0.36$ value \citep{piotrovich26}, for local AGNs $\sim 1.19$ \citep{piotrovich22}, for distant low luminosity AGNs (LLAGNs) $\sim 0.73$ \citep{piotrovich25b}. It can be assumed that this is most likely due to the selection bias.

\section{Technique for estimating physical parameters of AGNs}

\begin{figure}[ht!]
\centering
\includegraphics[bb= 30 10 715 535, clip, width=\linewidth]{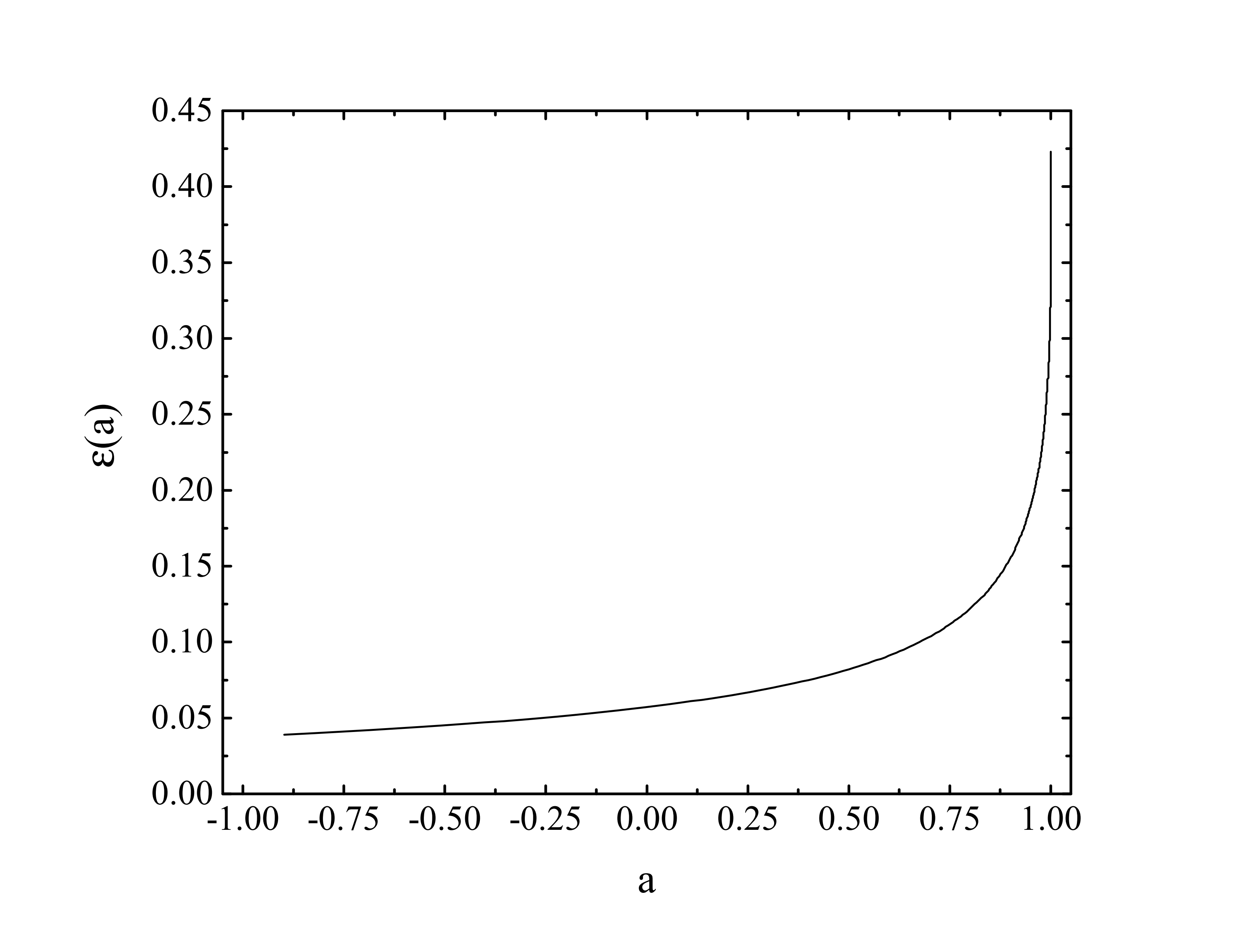}
\caption{Dependence of the radiative efficiency on the BH spin.
\label{fig:epsilon}}
\end{figure}

Here we, for self-consistency, used method similar to our previous works \citep{piotrovich22,piotrovich24,piotrovich25b}. Our method is, of course, highly model-dependent, but a more correct method for such a set of initial parameters has not yet been created. The fact that our method gives significantly different results for different types of AGN (and these results can be explained phenomenologically in terms of physical processes in AGN) can be considered as an indirect confirmation of the applicability of this method \citep{piotrovich22,piotrovich23,piotrovich25,piotrovich25b}.

The black hole spin can be estimated from the radiative efficiency of its accretion disk (see Fig.\ref{fig:epsilon}), $\varepsilon(a)$:
\begin{equation}
  \varepsilon = \frac{L_{\rm bol}}{\dot{M}c^2},
\end{equation}
where $\dot{M}$ is the accretion rate, $0.039 < \varepsilon < 0.324$ and $-1 \le a \le 0.998$ \citep{thorne74}. Negative $a$ corresponds to the retrograde rotation.

Using the Shakura--Sunyaev thin-disk model \citep{shakura73} we can estimate the radiative efficiency from the observable AGN parameters. Since our sample contains many objects with high redshift, we adopted statistical model from \citet{trakhtenbrot14} designed for distant objects (note that although \citet{trakhtenbrot14} considers AGNs with large masses, this method can also be used for low-mass objects):
\begin{equation}
\varepsilon = 0.073
\left(\frac{L_{\rm bol}}{10^{46}}\right)
\left(\frac{L_{\rm opt}}{10^{45}}\right)^{-1.5}
\left(\frac{4400\,{\rm \AA}}{5100\,{\rm \AA}}\right)^{-2}
M_8 \mu^{1.5},
\end{equation}

\noindent where $M_8 = M_{\rm BH}/10^8M_\odot$, $\mu=\cos i$, $i$ is an inclination angle between the line of sight and the normal to the accretion disk plane, $L_{\rm opt}$ is the optical luminosity (at 4400\,\AA). For self-consistency with our previous works we decided to use definition of optical luminosity $L_{\rm opt}$ from \citet{hopkins07}:
\begin{equation}
  \frac{L_\text{bol}}{L_\text{opt}} = 6.25 \left(\frac{L_\text{bol}}{10^{10} L_\odot}\right)^{-0.37} + 9.0 \left(\frac{L_\text{bol}}{10^{10} L_\odot}\right)^{-0.012},
\end{equation}

The angle $i$ was initially set to $45^\circ$ and then the angle value was changed up or down in 5 degree increments until model we were using produced a physically meaningful result. The SMBH mass for each object was recalculated to account for its dependence on $i$:
\begin{equation}
M_{\rm BH}=\frac{R_{\rm BLR}V_{\rm BLR}^2}{G}, \qquad
V_{\rm BLR}\simeq \frac{\mathrm{FWHM(H}\beta)}{2\sin i},
\end{equation}
where $R_{\rm BLR}$ was estimated following \citet{bentz13}:
\begin{equation}
  \log{\left(\frac{R_{\rm BLR}} {1\, \text{lt-day}}\right)} \approx 1.5 + 0.5\log{\left(\frac{L_{5100}} {10^{44} {\rm erg/s}}\right)}.
\end{equation}

The catalog from \citet{wu22} only provides values of $L_{5100}$ for the eight closest objects. For the remaining objects, we calculated these values from the bolometric luminosity, using a bolometric correction of $BC = 1.01$ \citep{richards06} for consistency with our previous work. These methods are certainly not precise, but we do not claim to accurately determine the parameters of specific objects; our study is primarily statistical in nature.

The method described above uses the FWHM(H$\beta$) as an example. However, this conversion method is actually independent of the spectral line used to determine the mass, only the change in the angle $i$ affects the final result.

The spin was then determined numerically using method from \citep{bardeen72}:
\begin{equation}
\varepsilon(a)=1-\frac{R_{\rm ISCO}^{3/2}-2R_{\rm ISCO}^{1/2}+|a|}
{R_{\rm ISCO}^{3/4}(R_{\rm ISCO}^{3/2}-3R_{\rm ISCO}^{1/2}+2|a|)^{1/2}},
\end{equation}
where $R_{\rm ISCO}$ is the radius of the innermost stable circular orbit:
\begin{equation}
  \begin{array}{l}
   R_{\rm ISCO}(a) = 3 + Z_2 \pm [(3 - Z_1)(3 + Z_1 + 2 Z_2)]^{1/2},\\
   Z_1 = 1 + (1 - a^2)^{1/3}\left[(1 + a)^{1/3} + (1 - a)^{1/3}\right],\\
   Z_2 = (3 a^2 + Z_1^2)^{1/2}.
  \end{array}
\end{equation}
Here ''-'' is used when $a \geq 0$, and ''+'' when $a < 0$.

\section{Analysis of the estimated spin values}

\renewcommand{\arraystretch}{0.2}
\begin{table*}
\begin{center}
\caption[]{Table shows object name, cosmological redshift $z$, bolometric luminosity $\log(L_{\rm bol}{\rm [erg/s]})$, SMBH mass $\log(M_{\rm BH}/M_\odot)$) Eddington ratio $\log(l_{\rm E})$ and results of our estimations: inclination angle $i$ (in degrees), radiative efficiency $\varepsilon$, spin value $a$, new SMBH mass $\log(M^*_{\rm BH}/M_\odot)$ and new Eddington ratio $\log(l^{*}_{\rm E})$.}
\begin{ruledtabular}
\begin{tabular}{lccccccccc}
SDSS name & $z$ & $L_{\rm bol}$ & $M_{\rm BH}$ & $l_{\rm E}$ & $i$ & $\varepsilon$ & $a$ & $M^*_{\rm BH}$ & $l^*_{\rm E}$ \\
\noalign{\smallskip}
\hline
000605.17-095611.3 & 1.673000 & 41.03 &  5.34 & -2.42 & 45 & 0.248 &  0.986 &  4.93 & -2.01 \\
002526.90+135122.8 & 2.143000 & 43.41 &  6.22 & -0.92 & 30 & 0.042 & -0.624 &  6.11 & -0.81 \\
003410.78-010030.5 & 1.904956 & 41.32 &  6.17 & -2.97 & 65 & 0.249 &  0.988 &  5.54 & -2.34 \\
003423.52-001004.5 & 1.955430 & 41.10 &  6.28 & -3.29 & 75 & 0.216 &  0.972 &  5.60 & -2.60 \\
004502.43-040550.7 & 1.914000 & 40.91 &  5.80 & -3.00 & 65 & 0.261 &  0.990 &  5.17 & -2.37 \\
010036.99-011305.3 & 1.803729 & 40.56 &  4.78 & -2.34 & 45 & 0.200 &  0.960 &  4.37 & -1.92 \\
011810.17+202535.0 & 2.109000 & 41.79 &  5.80 & -2.12 & 45 & 0.134 &  0.848 &  5.39 & -1.70 \\
014245.01+140348.0 & 1.794393 & 40.71 &  5.17 & -2.58 & 50 & 0.256 &  0.990 &  4.69 & -2.09 \\
014503.02+094312.2 & 1.676000 & 40.21 &  5.36 & -3.27 & 75 & 0.201 &  0.962 &  4.68 & -2.57 \\
020016.95+133745.0 & 0.064029 & 41.72 &  5.82 & -2.21 & 45 & 0.164 &  0.918 &  5.41 & -1.80 \\
020814.99+071634.3 & 1.017014 & 40.13 &  4.20 & -2.18 & 45 & 0.141 &  0.868 &  3.79 & -1.77 \\
073856.16+184937.7 & 0.113555 & 41.89 &  5.67 & -1.90 & 45 & 0.082 &  0.512 &  5.26 & -1.48 \\
074109.38+270432.9 & 1.972710 & 42.02 &  6.28 & -2.37 & 45 & 0.251 &  0.988 &  5.87 & -1.95 \\
075044.94+222430.5 & 0.840064 & 39.57 &  4.57 & -3.12 & 70 & 0.235 &  0.982 &  3.91 & -2.45 \\
084757.01+303239.1 & 1.864230 & 41.10 &  5.72 & -2.74 & 55 & 0.278 &  0.994 &  5.18 & -2.19 \\
084857.84+232207.5 & 1.511000 & 41.13 &  5.80 & -2.79 & 55 & 0.314 &  0.998 &  5.26 & -2.24 \\
085048.04+034231.7 & 1.765000 & 41.29 &  5.50 & -2.32 & 45 & 0.200 &  0.960 &  5.09 & -1.90 \\
091102.47+314309.1 & 1.872380 & 41.75 &  6.08 & -2.45 & 45 & 0.282 &  0.994 &  5.67 & -2.03 \\
092605.34+382308.1 & 1.680000 & 40.85 &  5.90 & -3.16 & 70 & 0.258 &  0.990 &  5.24 & -2.50 \\
093511.55+001331.6 & 1.739000 & 41.19 &  5.95 & -2.87 & 60 & 0.280 &  0.994 &  5.36 & -2.28 \\
093852.98-021520.0 & 1.875000 & 41.06 &  5.79 & -2.85 & 60 & 0.260 &  0.990 &  5.20 & -2.25 \\
095427.86+101053.5 & 0.037561 & 40.06 &  5.35 & -3.40 & 75 & 0.276 &  0.994 &  4.67 & -2.71 \\
102841.46+445749.4 & 1.798692 & 41.30 &  5.23 & -2.04 & 45 & 0.104 &  0.710 &  4.82 & -1.62 \\
103953.46+265233.4 & 1.590000 & 40.96 &  5.94 & -3.09 & 70 & 0.218 &  0.974 &  5.28 & -2.43 \\
111812.63+335031.6 & 1.734086 & 40.14 &  5.11 & -3.09 & 65 & 0.319 &  0.998 &  4.48 & -2.46 \\
113953.54+072905.8 & 1.640000 & 41.19 &  5.53 & -2.46 & 45 & 0.269 &  0.992 &  5.12 & -2.04 \\
122212.66+461321.2 & 1.588405 & 39.95 &  5.99 & -4.16 & 85 & 0.292 &  0.996 &  5.28 & -3.44 \\
122509.16+061319.3 & 1.916000 & 41.09 &  5.96 & -2.98 & 65 & 0.254 &  0.988 &  5.33 & -2.35 \\
124816.23+400223.4 & 1.399385 & 40.20 &  6.14 & -4.06 & 85 & 0.228 &  0.978 &  5.43 & -3.34 \\
131704.56+483755.1 & 0.439623 & 42.43 &  5.93 & -1.62 & 45 & 0.049 & -0.262 &  5.52 & -1.19 \\
134135.19+361911.1 & 1.984000 & 41.53 &  5.06 & -1.65 & 45 & 0.043 & -0.564 &  4.65 & -1.23 \\
135617.27+634124.9 & 0.916085 & 39.83 &  4.74 & -3.02 & 65 & 0.282 &  0.994 &  4.11 & -2.40 \\
141853.75+564023.5 & 1.837261 & 40.87 &  5.95 & -3.20 & 70 & 0.274 &  0.994 &  5.29 & -2.53 \\
142934.91+650815.9 & 1.983000 & 42.11 &  5.71 & -1.71 & 45 & 0.056 & -0.004 &  5.30 & -1.29 \\
143051.48+365109.0 & 1.910030 & 41.05 &  4.85 & -1.91 & 45 & 0.077 &  0.446 &  4.44 & -1.50 \\
145856.85+594058.8 & 1.588560 & 40.34 &  4.44 & -2.21 & 45 & 0.150 &  0.890 &  4.03 & -1.79 \\
152801.84+340942.9 & 0.834000 & 42.08 &  6.01 & -2.04 & 45 & 0.118 &  0.788 &  5.60 & -1.62 \\
153027.62+351315.0 & 0.126477 & 43.15 &  6.30 & -1.26 & 40 & 0.040 & -0.754 &  5.97 & -0.93 \\
160705.98+465117.7 & 1.874000 & 39.77 &  5.28 & -3.62 & 80 & 0.247 &  0.986 &  4.58 & -2.92 \\
170054.18+413113.3 & 1.783000 & 40.17 &  5.23 & -3.18 & 70 & 0.264 &  0.992 &  4.57 & -2.51 \\
212109.84+005902.5 & 0.893836 & 40.29 &  5.14 & -2.96 & 65 & 0.237 &  0.982 &  4.51 & -2.33 \\
212801.52+042209.2 & 5.807989 & 41.77 &  5.63 & -1.98 & 45 & 0.095 &  0.644 &  5.22 & -1.56 \\
212908.71-005733.1 & 1.598000 & 39.51 &  4.90 & -3.51 & 80 & 0.193 &  0.954 &  4.20 & -2.80 \\
213441.61+005228.2 & 0.183846 & 42.84 &  6.14 & -1.41 & 40 & 0.049 & -0.262 &  5.81 & -1.08 \\
214318.54-011717.8 & 1.695856 & 40.62 &  5.12 & -2.61 & 50 & 0.278 &  0.994 &  4.64 & -2.13 \\
214353.43-004649.2 & 1.815000 & 40.68 &  5.21 & -2.64 & 50 & 0.299 &  0.998 &  4.73 & -2.16 \\
214548.31+011209.2 & 1.827436 & 40.96 &  4.90 & -2.05 & 45 & 0.104 &  0.710 &  4.49 & -1.63 \\
215355.65+000913.5 & 1.907000 & 41.37 &  5.19 & -1.93 & 45 & 0.082 &  0.512 &  4.78 & -1.51 \\
220340.86-000303.3 & 1.986345 & 41.51 &  5.65 & -2.26 & 45 & 0.176 &  0.936 &  5.24 & -1.84 \\
222606.25+262639.0 & 2.028263 & 43.81 &  6.16 & -0.46 & 20 & 0.045 & -0.452 &  6.38 & -0.68 \\
223746.95+275736.4 & 0.844249 & 40.03 &  5.76 & -3.84 & 85 & 0.141 &  0.868 &  5.05 & -3.13 \\
224026.90+303000.6 & 0.890728 & 39.01 &  4.49 & -3.60 & 80 & 0.247 &  0.986 &  3.79 & -2.89 \\
225429.64+033652.0 & 1.644031 & 43.48 &  6.28 & -0.91 & 30 & 0.042 & -0.624 &  6.17 & -0.79 \\
230626.32+274301.5 & 1.909000 & 40.99 &  5.77 & -2.89 & 60 & 0.290 &  0.996 &  5.18 & -2.30 \\
231346.28+260008.2 & 0.104310 & 41.65 &  5.55 & -2.02 & 45 & 0.104 &  0.710 &  5.14 & -1.60 \\
232939.35+104701.5 & 1.646997 & 40.46 &  5.71 & -3.36 & 75 & 0.253 &  0.988 &  5.03 & -2.68 \\
235217.85+115613.2 & 0.079067 & 42.33 &  5.94 & -1.72 & 45 & 0.061 &  0.136 &  5.53 & -1.30 \\
235708.79-090028.0 & 1.789000 & 41.21 &  5.69 & -2.59 & 50 & 0.276 &  0.994 &  5.21 & -2.11 \\
\end{tabular}
\end{ruledtabular}
\label{table_01}
\end{center}
\end{table*}
\renewcommand{\arraystretch}{1.0}

\begin{figure}[ht!]
\includegraphics[bb= 30 5 715 535, clip, width=\linewidth]{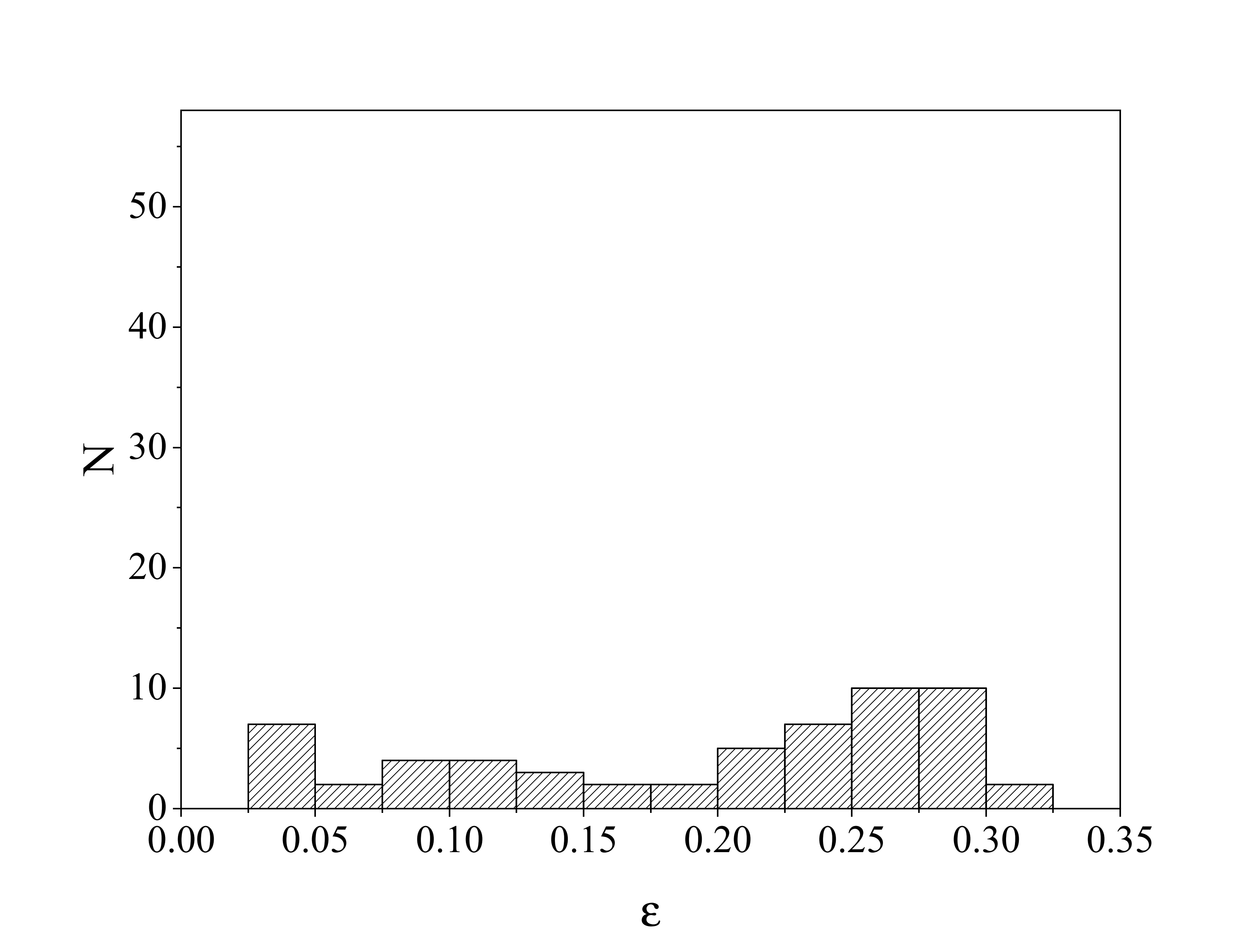}
\caption{Distribution of the estimated radiative efficiency.
\label{fig:hist_eps}}
\end{figure}

\begin{figure}[ht!]
\includegraphics[bb= 30 10 715 535, clip, width=\linewidth]{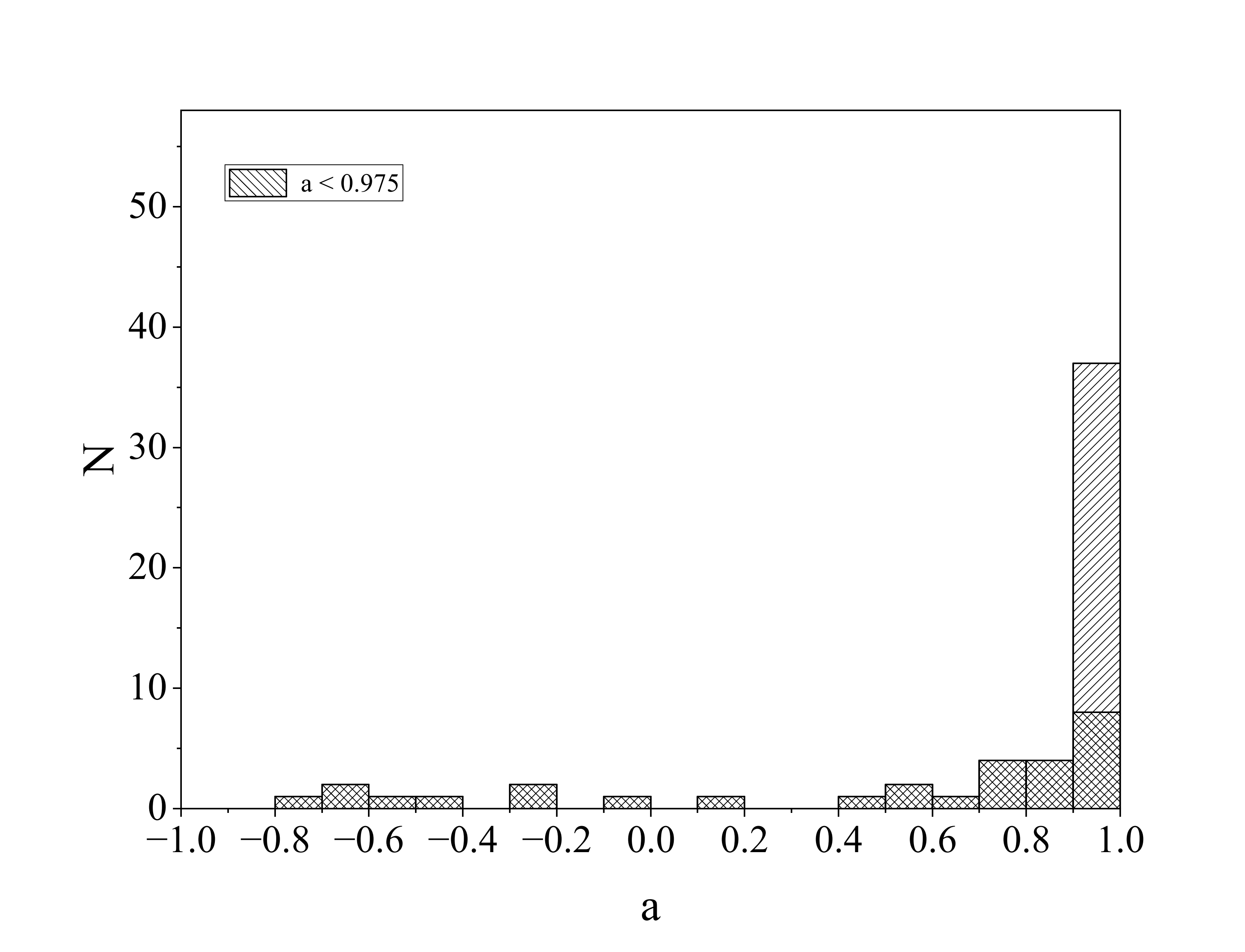}
\caption{Distribution of the estimated spin values.
\label{fig:hist_a}}
\end{figure}

\begin{figure}[ht!]
\includegraphics[bb= 30 5 715 535, clip, width=\linewidth]{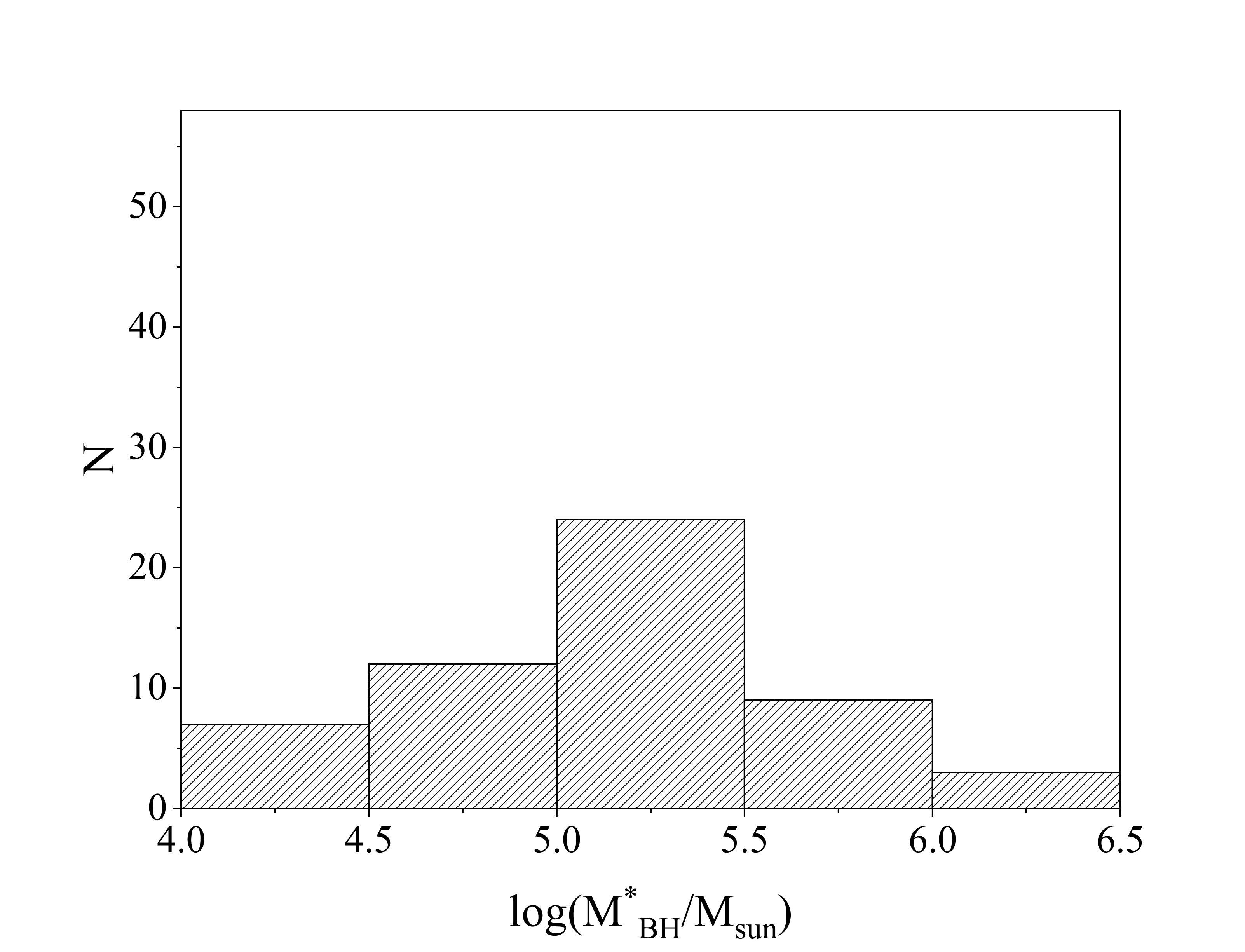}
\caption{Distribution of the estimated SMBH masses.
\label{fig:new_hist_MBH}}
\end{figure}

\begin{figure}[ht!]
\includegraphics[bb= 30 10 715 535, clip, width=\linewidth]{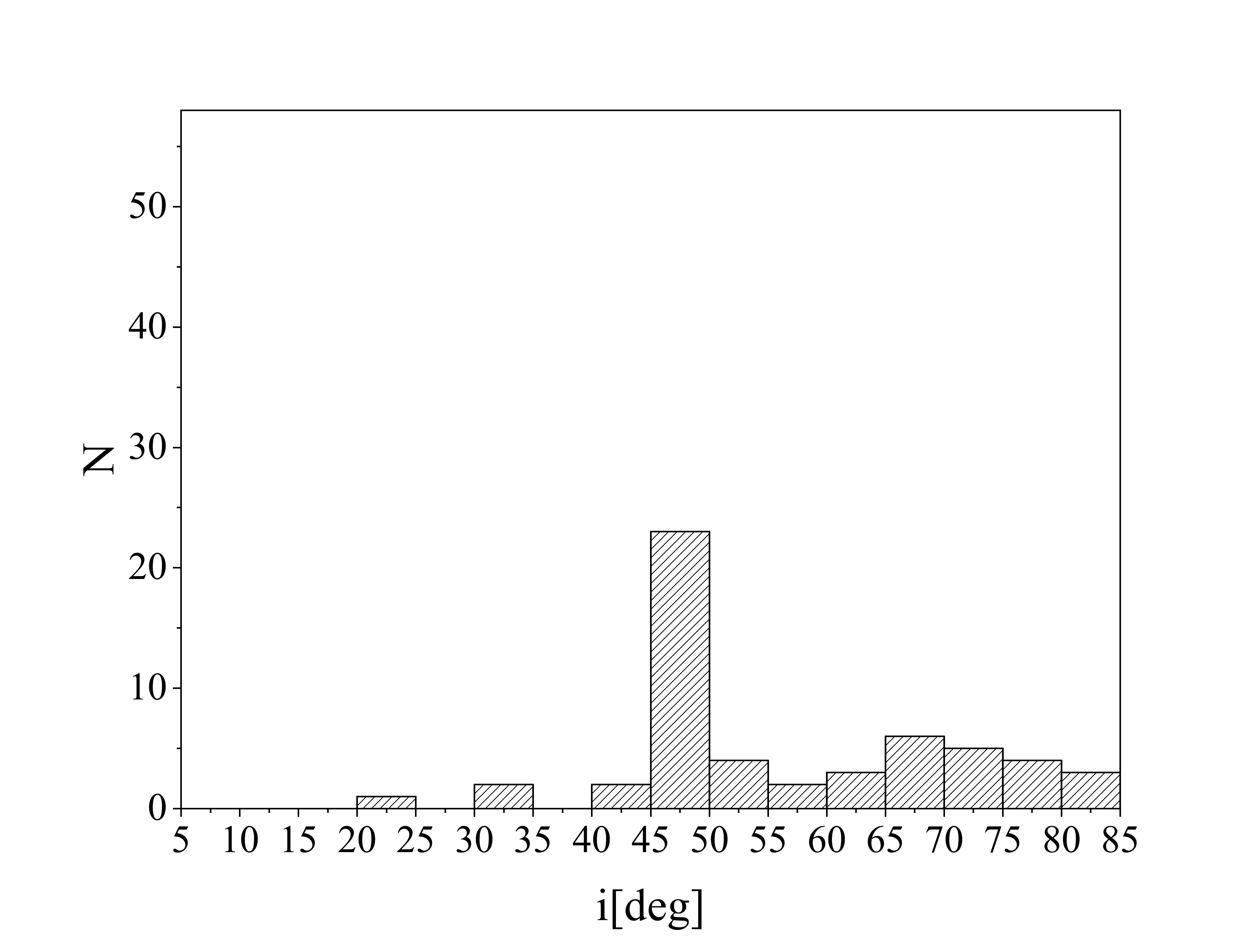}
\caption{Distribution of the estimated inclination angles.
\label{fig:hist_i}}
\end{figure}

We estimated the spins of the objects in our sample using the method described above. Table \ref{table_01} presents our results. It can be seen that 8 objects show signs of retrograde rotation ($a < 0$). It should be noted that we do not claim to accurately determine the spin values for each specific object (we do not have enough observational data for this) our estimations are primarily of qualitative, statistical significance.

One can see that many objects have inclination angle values that exceed 60 degrees. In the case of high-mass AGNs in large galaxies, the presence of a dust torus significantly affects the spectra of objects at inclination angles greater than 60 degrees. However, the presence or absence of a dust torus in dwarf galaxies remains an open and unresolved question. In this paper, we assume that, due to the low accretion rate and low metallicity, little dust is produced, and dust torus formation is virtually nonexistent. And observed infrared excess in some objects of this type is explained by high star formation rate in dwarf galaxies \citep{hirashita99,elitzur06,elitzur09,cao26}. To further clarify the issue, we will conduct an additional statistical analysis of our set of objects, excluding objects with an inclination angle greater than 60 degrees.

The following is a statistical analysis of the results obtained.

Figs.\ref{fig:hist_eps}-\ref{fig:hist_i} shows the distribution of the estimated radiative efficiency, the spin values, SMBH masses and inclination angles. One can see that the mass distribution has a similar log-normal shape, but the peak has shifted to the lower mass region. This is due to the fact that we recalculated the masses to account for the new inclination angle values, and, as can be seen from the angle distribution, a significant portion of the angles fall in the range greater than 45 degrees, at which point the calculated mass value decreases. Also important is the fact that \citet{wu22} used the method from \citet{vestergaard06} (with a constant tilt angle of $\sim 35$ degrees) to determine the mass, while we used a more modern and more accurate method from \citet{bentz13}. The distribution of radiative efficiency appears relatively uniform and does not show any pronounced features. The spin distribution appears to be fairly typical for AGNs \citep{trakhtenbrot14,afanasiev18,piotrovich22,piotrovich24,piotrovich25b,daly19,reynolds21,azadi23}. $\sim 64$\% of all objects are within $a > 0.9$.

\begin{figure}[ht!]
\centering
\includegraphics[bb= 30 5 715 535, clip, width=\linewidth]{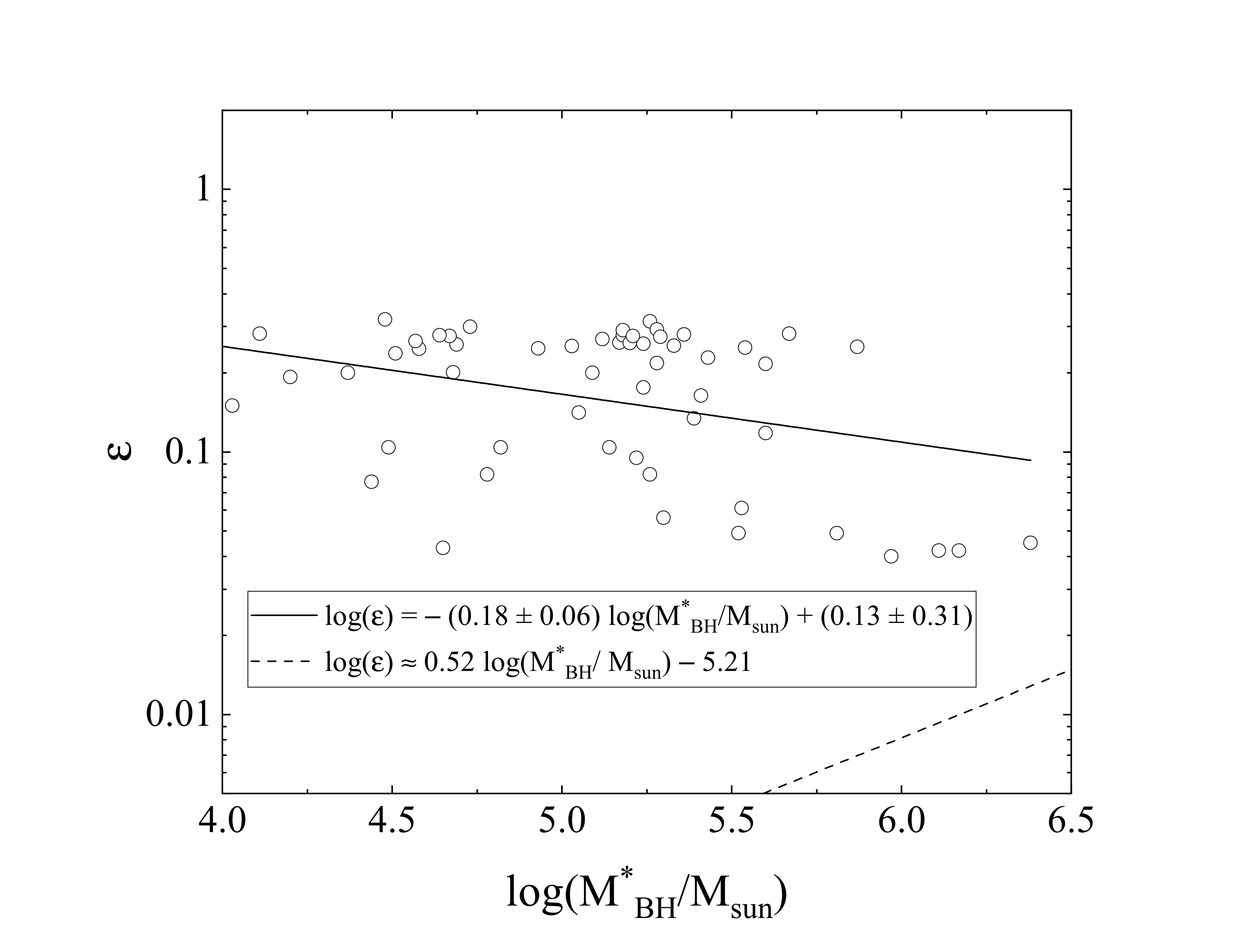}
\caption{Dependency of the estimated radiative efficiency on SMBH mass. The dashed line is the linear fit for the sample from \citet{davis11}.
\label{fig:eps_MBH}}
\end{figure}

Fig.\ref{fig:eps_MBH} presents the dependency of the estimated radiative efficiency on SMBH mass. A moderate anticorrelation is observed between radiative efficiency and SMBH mass (Pearson correlation coefficient is -0.38). Linear fitting gives us:
\begin{equation}
\log(\varepsilon) = - (0.18 \pm 0.06) \log(M^*_{\rm BH} / M_\odot) + (0.13 \pm 0.31).
\label{eq:eps_MBH}
\end{equation}
This expression should not be seen as an exact correlation, but rather as a manifestation of a certain trend. Note that a similar relationship obtained by us, for example, for red quasars \citep{piotrovich26}, and obtained in \citet{davis11} for their AGN sample, has a slope of $\sim 0.5$. Thus, it can be seen that for low-mass AGN this dependence is qualitatively different and has the opposite sign.

\begin{figure}[ht!]
\centering
\includegraphics[bb= 30 5 715 535, clip, width=\linewidth]{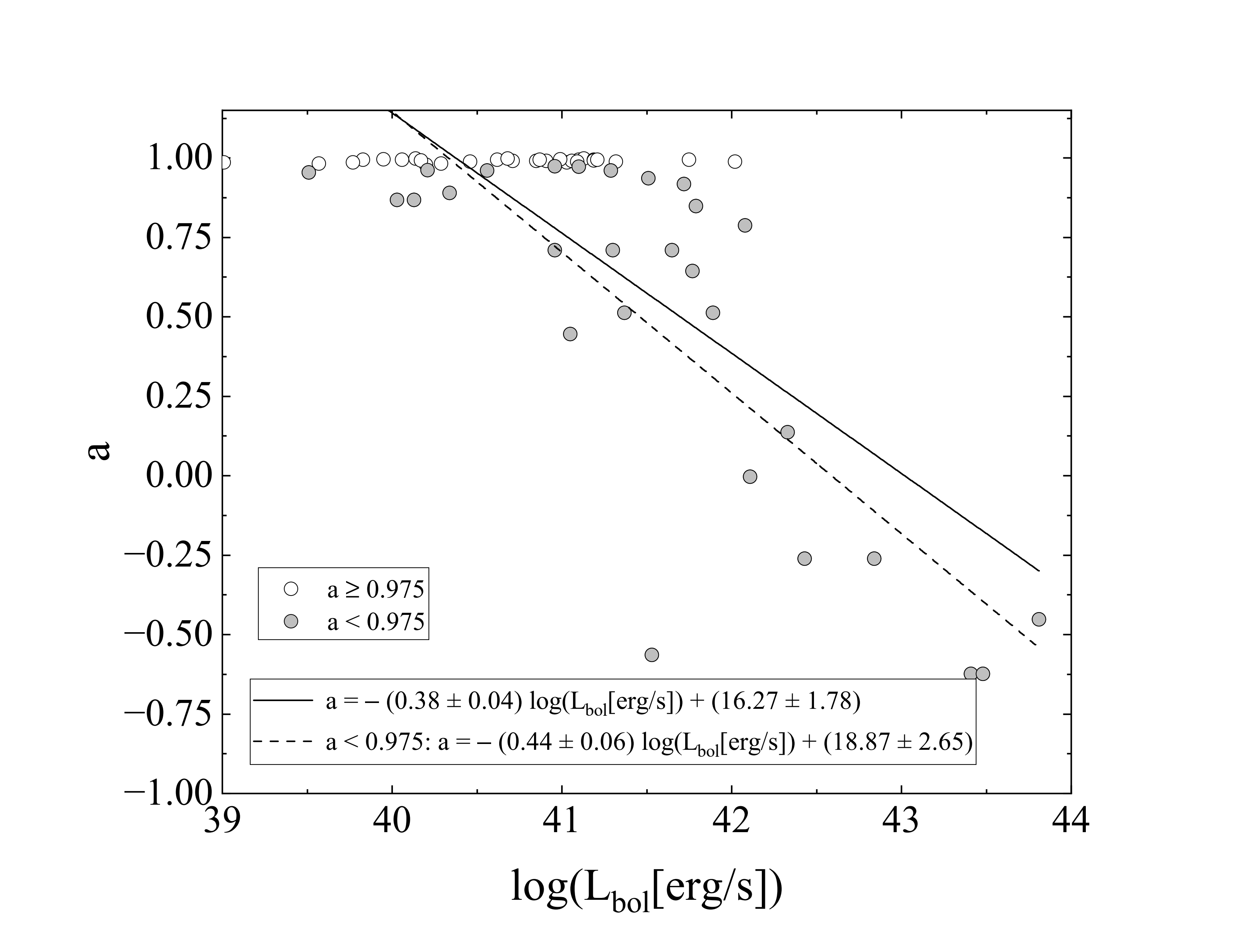}
\caption{Dependency of the estimated spin values on bolometric luminosity.
\label{fig:a_Lbol}}
\end{figure}

\begin{figure}[ht!]
\centering
\includegraphics[bb= 30 10 715 535, clip, width=\linewidth]{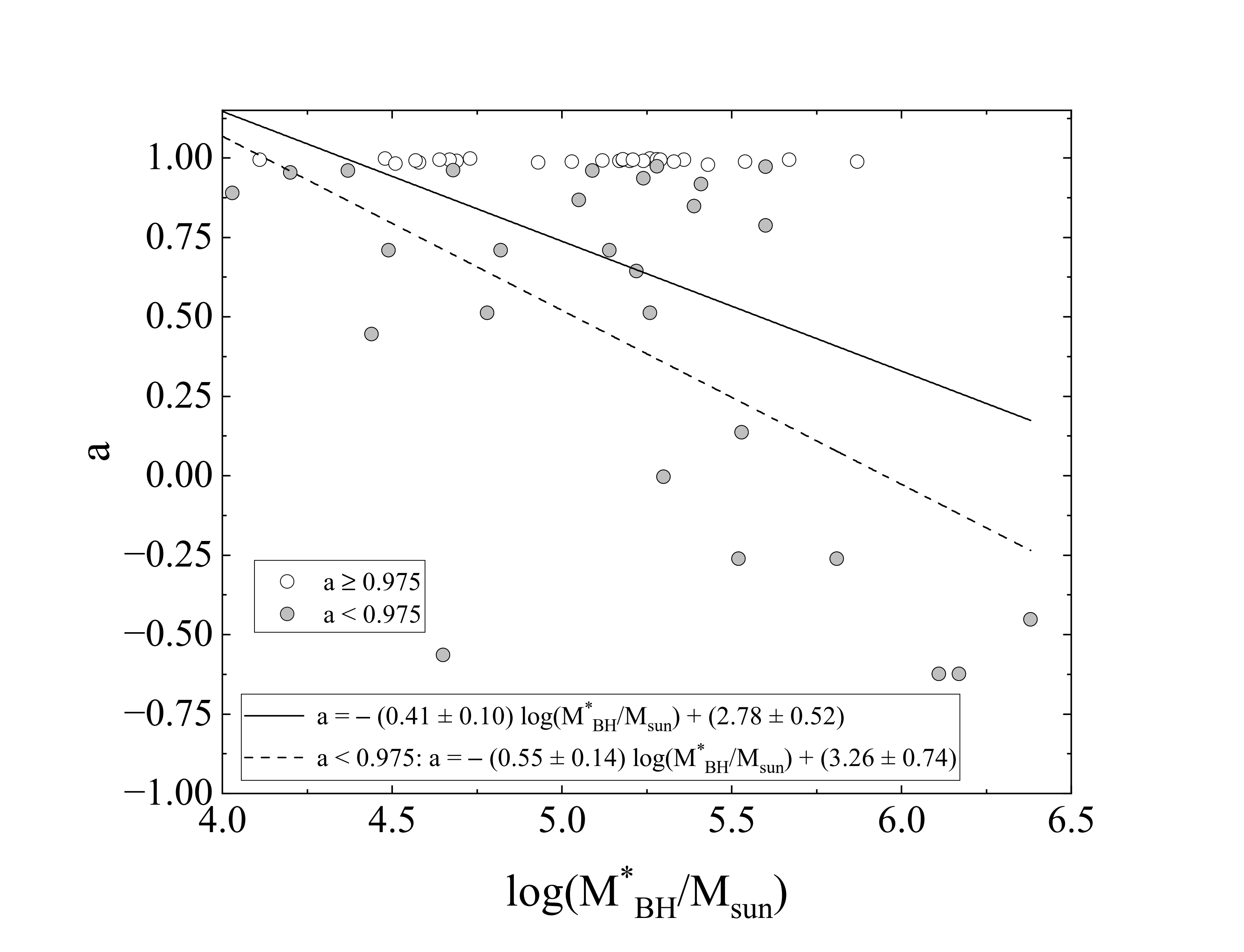}
\caption{Dependency of the estimated spin values on SMBH mass.
\label{fig:a_MBH}}
\end{figure}

\begin{figure}[ht!]
\centering
\includegraphics[bb= 30 5 715 540, clip, width=\linewidth]{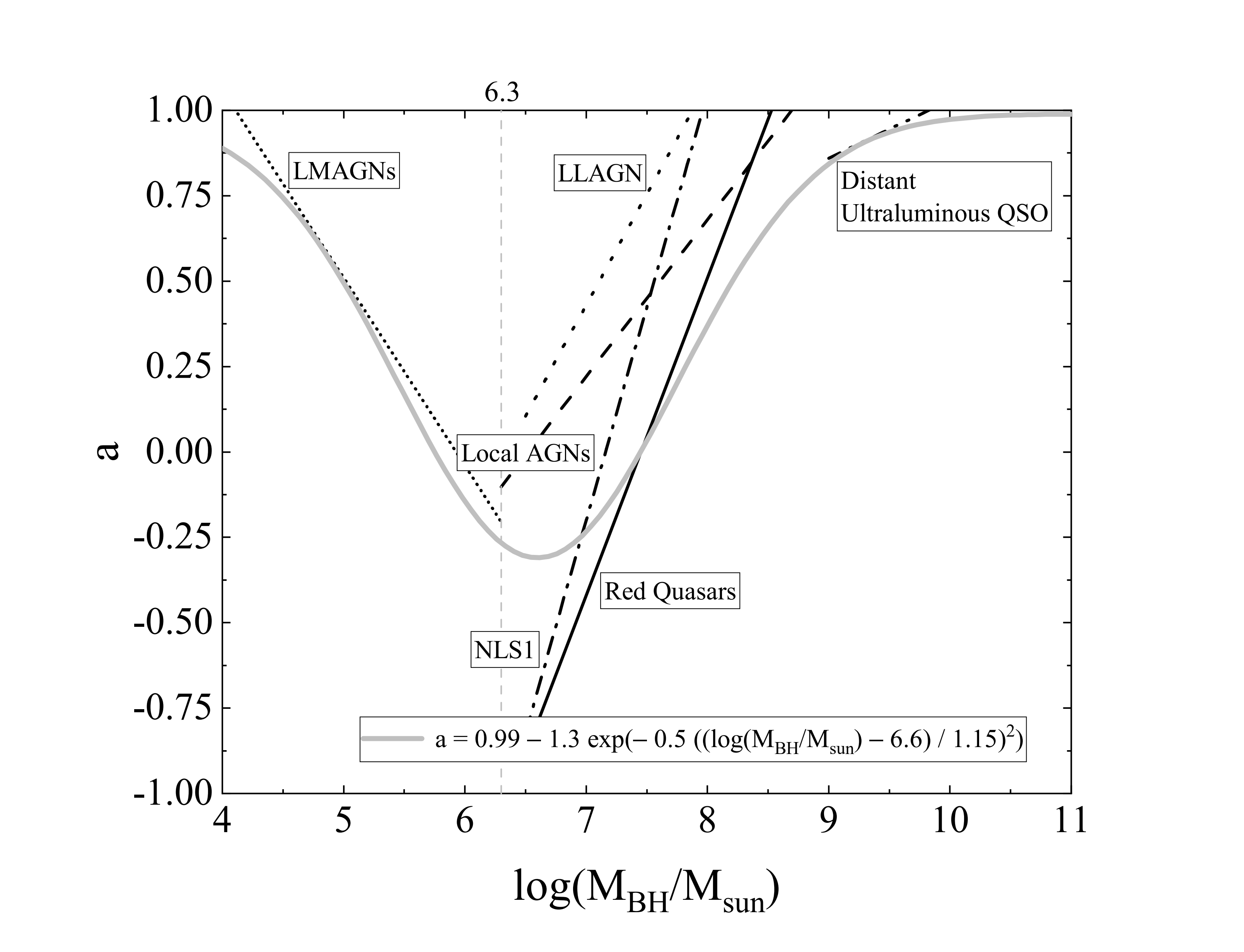}
\caption{The spin--mass dependence linear fits for different types of objects.
\label{fig:fits}}
\end{figure}

\begin{table*}
\begin{center}
\caption[]{The spin--mass dependence linear fits for different types of objects. LLAGN is Low Luminosity AGN.}
\begin{ruledtabular}
\begin{tabular}{lll}
Object type                   & Linear fit                                                         &  Reference\\
\hline
Distant Ultraluminous Quasars &  $a = (0.17 \pm 0.05) \log(M_{\rm BH}/M_\odot) - (0.67 \pm 0.49)$  &  \citet{piotrovich25}\\
Local AGNs                    &  $a = (0.46 \pm 0.09) \log(M_{\rm BH}/M_\odot) - (3.00 \pm 0.71)$  &  \citet{piotrovich22}\\
LLAGN                         &  $a = (0.65 \pm 0.11) \log(M_{\rm BH}/M_\odot) - (4.12 \pm 0.82)$  &  \citet{piotrovich25b}\\
Red Quasars                   &  $a = (0.93 \pm 0.03) \log(M_{\rm BH}/M_\odot) - (6.93 \pm 0.26)$  &  \citet{piotrovich26} \\
NLS1                          &  $a = (1.25 \pm 0.05) \log(M_{\rm BH}/M_\odot) - (8.95 \pm 0.35)$  &  \citet{piotrovich23}\\
LMAGNs                        &  $a = -(0.55 \pm 0.14) \log(M_{\rm BH}/M_\odot) + (3.26 \pm 0.74)$ &  Present work\\
\end{tabular}
\end{ruledtabular}
\label{tab:fits}
\end{center}
\end{table*}

Figures \ref{fig:a_Lbol} and \ref{fig:a_MBH} show the dependencies of estimated spin on bolometric luminosity and SMBH mass. To take into account the saturation effect at high spin values, we also examined the properties of a sample of only objects with $a < 0.975$. In the Fig.\ref{fig:hist_a} one can see that for such objects the distribution of spins becomes significantly more uniform. For spin--luminosity dependence one can see strong anticorrelation between parameters (Pearson correlation coefficients are -0.76 for all objects and -0.80 for objects with $a < 0.975$). For spin--mass dependence moderate to strong anticorrelation between parameters is observed (Pearson correlation coefficients are -0.47 for all objects and -0.59 for objects with $a < 0.975$). To take into account the possible effect of the redshift distribution not being completely normal (see Fig.\ref{fig:hist_z}), we also consider spin--mass relation only for objects with $a < 0.975$ and $1.0 \leq z \leq 2.5$ (for this set Pearson correlation coefficients is -0.58) and for objects with $i < 60^\circ$ (Pearson's $r = -0.50$). Linear fitting gives us:
\begin{equation}
\begin{aligned}
&a = - (0.38 \pm 0.04) \log(L_{\rm bol}{\rm [erg/s]}) + (16.27 \pm 1.78),\\
&a < 0.975:\\
&a = - (0.44 \pm 0.06) \log(L_{\rm bol}{\rm [erg/s]}) + (18.87 \pm 2.65);\\
\end{aligned}
\label{eq:a_Lbol}
\end{equation}
\begin{equation}
\begin{aligned}
&a = - (0.41 \pm 0.10) \log(M^*_{\rm BH}/ M_\odot) + (2.78 \pm 0.52),\\
&a < 0.975:\\
&a = - (0.55 \pm 0.14) \log(M^*_{\rm BH}/ M_\odot) + (3.26 \pm 0.74),\\
&a < 0.975,\, 1.0 \leq z \leq 2.5:\\
&a = - (0.53 \pm 0.19) \log(M^*_{\rm BH}/ M_\odot) + (3.13 \pm 0.95).\\
&i < 60^{\circ}:\\
&a = - (0.51 \pm 0.15) \log(M^*_{\rm BH}/ M_\odot) + (3.19 \pm 0.77).\\
\end{aligned}
\label{eq:a_MBH}
\end{equation}

As for the large apparent errors in the intercept values of linear fittings, this is not due to the data itself, but to the specific fitting method. For example, if, instead of the $L_{\rm bol}$[erg/s], we take $L_{\rm bol}/10^{41} {\rm erg/s}$ (i.e. divided by mean value), the fitting described above will look like this: $a = - (0.38 \pm 0.04) \log(L_{\rm bol} / 10^{41}{\rm erg/s}) + (0.76 \pm 0.04)$. The same applies to the dependence of spin on mass. However, we used this representation in our previous works and in this paper we plot on the same graph both the linear approximation obtained in this work and the approximations we obtained earlier in previous studies. To do this, they must all be obtained using the same method.

In the spin--luminosity dependence the saturation effect does not play a significant role, but in the spin--mass dependence it has a noticeable effect, albeit within the error limits. It can be seen that the properties of objects with an inclination angle $i < 60^{\circ}$ differ slightly, within the error limits, from all objects in the sample, but are very close to objects with $a < 0.975$. It can also be seen that the redshift distribution does not have a noticeable effect on the result. Therefore, in the case of the spin--mass dependence, we decided to use a value for all objects from our sample with $a < 0.975$. Note that our result is thus actually almost independent of whether the dwarf galaxies have dust tori. This type of spin--mass dependence is very different from similar ones for other types of AGN (see Table \ref{tab:fits}). This result can be compared with the results obtained by a completely different method by \citet{mallick22}. Their study contains significantly fewer low-mass AGNs (13), and they are located in a narrower mass range ($5.5 < \log(M_{\rm BH} / M_\odot) < 6.5$). However, their general conclusion that the spins for objects in this mass range are, on average, lower than for massive AGNs is consistent with ours.

This may indicate that for our sample of objects, the main mechanisms of mass growth are mergers and/or chaotic accretion, which on average reduce spin, rather than disk accretion, which effectively increases it. This conclusion contradicts current ideas, which suggest disk accretion as the primary mechanism. This contradiction could, in principle, be explained by a selection bias, as these objects are quite dim and are best seen when surrounded by little matter, leading to a comparatively low accretion rate (the low Eddington ratios also support this), making mergers and/or chaotic accretion the primary mechanisms of mass growth.

On the other hand, if we assume that this is not a selection bias, but a real property of most objects of a this type, then we can advance the following working hypothesis. Initially, low-mass SMBHs are formed by the merger of intermediate-mass black holes. This merger mechanism should produce objects with initially high spin values. Moreover, such objects will be quite numerous (we see very few of them because they are distant and dim and therefore very difficult to observe), so the merger rate will be quite high. Also, since accretion disks either have not yet formed or have been destroyed (or deformed) as a result of mergers, chaotic accretion will play a significant role. And this two mechanisms will be the primary mechanisms for mass growth. Subsequently, as the number of objects becomes smaller and the distances between them grow, the merger rate will decrease, and at some point, disk accretion becomes the primary mechanism for mass growth. Later on, due to the depletion of matter in the central regions of the host galaxies, the accretion rate begins to decrease, and mergers may again begin to play a significant role in mass growth. An illustration of this hypothesis is shown in the Fig.\ref{fig:fits}. The gray curve ($a \approx 0.99 - 1.3 \exp(-0.5 ((\log(M_{\rm BH}/ M_\odot) - 6.6)/1.15)^2)$) shows the general form of the hypothetical spin--mass relationship for SMBHs of all types. We emphasize that this is, of course, not an exact result, but only a highly simplified, hypothetical and approximate relationship obtained through fitting.

This result can be compared with that of \citet{krause25} (Figure 5 from their paper). At first glance, their result seems to be the complete opposite of ours. However, it's worth noting that their work uses tens of times fewer objects than our studies on this topic and the objects are located in a narrower mass range. Regarding large-mass AGNs, some of our studies used samples of about a thousand objects \citep{piotrovich23,piotrovich26}. Therefore, the lack of statistical data in \citet{krause25} prevents us from conclusively stating that our results contradict theirs. In fact, it's possible that they even agree.

\section{Conclusions}

We estimated the spins of a sample of 58 low-mass AGNs. Analysis of the obtained spins showed that they decrease with increasing SMBH mass, leading us to hypothesize that (if we assume that this is a real effect and not selection bias) mergers and/or chaotic accretion are the primary mechanisms for mass growth in these objects. This, in turn, allowed us to propose a more general hypothesis about the evolution of AGNs in general. The essence of this hypothesis is that early low-mass SMBHs, which are formed by the mergers of intermediate-mass black holes, have high spins. Then, during their evolution, the spins, on average, initially decrease and then begin to increase, with the rate of increase gradually slowing (see Fig.\ref{fig:fits}).

\begin{acknowledgments}
We are grateful to the Reviewer for very useful comments.
\end{acknowledgments}

\end{document}